\documentstyle[preprint,revtex]{aps}
\begin{document}
\rightline{McGill/92-07}
\rightline{{\tt hep-ph/9207242}}
\rightline{March 1992}
\bigskip
\bigskip
\begin{title}
Are There Oscillations In The Baryon/Meson Ratio?
\end{title}
\bigskip
\author{Jean-Ren\'e Cudell\cite{compaddress}  and\
    Keith R.~Dienes\cite{compaddress2}}
\bigskip
\begin{instit}
Dept.~of Physics, McGill University
3600 University St., Montr\'eal, Qu\'ebec, Canada~~ H3A-2T8
\end{instit}
\bigskip
\bigskip
\begin{abstract}
All available data indicate a surplus of
baryon states over meson states for energies greater than about 1.5 GeV.
Since hadron-scale string theory suggests that
their numbers should become equal with increasing
energy, it has recently been proposed that there must exist exotic
mesons with masses just above 1.7 GeV in order to fill the
deficit.  We demonstrate that a string-like
picture is actually consistent with the present numbers
of baryon and meson states, and in fact predicts regular
oscillations in their ratio.  This suggests a different
role for new hadronic states.
\end{abstract}
\smallskip
\newpage
\narrowtext


In a recent work, Freund and Rosner\cite{freuros} have examined
the separate densities of observed meson and baryon states as functions
of their masses.  They find that the integrated number of baryon
states is less than that of meson states for masses less
than about 1.7 GeV, but then greatly surpasses the
meson number at higher energies.  Since hadron-scale string theories are
successful in modelling not only the hadronic Regge
trajectories but also the exponential (Hagedorn) growth\cite{hagref}
in the total hadronic density, Freund and Rosner point out that such
theories may also serve as the basis for understanding
the relation {\it between} the separate meson and baryon densities.
This is possible in part
due to a recent result of Kutasov and Seiberg\cite{kutsei} which
states that the numbers of bosonic and fermionic states in a
non-supersymmetric tachyon-free string theory must approach each
other as increasingly massive states are included.
On the basis of this theoretical result, Freund and Rosner predict
that there must exist a number of mesons yet to be
discovered with masses above 1.7 GeV (in order to match the rise
in baryon number); furthermore, since the presently-observed
baryon/meson ratio is consistent with quark-model calculations
which include only conventional mesons and baryons\cite{quarkmodel}
({\it i.e.}, states with $q\overline{q}$ and $qqq$ quark configurations
respectively), they additionally speculate that these new mesons are likely
to be exotic (with quark content $q^{p+1}\, \overline{q}^{p+1}$, $p\geq 1$).
This then implies the existence of exotic baryons (with
configurations $q^{p+3}\, \overline{q}^p$, $p\geq 1$), and
one is led to imagine a tower of exotic hadronic
states with higher and higher masses.

In this letter we first present a more refined
analysis of the existing data and then examine more precisely the role a
hadron-scale string theory might play in predicting the densities
of baryon and meson states.  In particular, while the result of
Kutasov and Seiberg can be expected to hold in the {\it asymptotic}
region (mass $M\to \infty$), we find that for energies in the
GeV range
a na\"\i ve hadron-scale string picture implies that the ratio between
the numbers of baryon and meson states should in fact {\it oscillate}
around unity, with mesons favored first, then baryons, then mesons again.
The amplitude of this oscillation falls to zero as the mass increases
(in accordance with the Kutasov-Seiberg result), but we find that
for masses below 2 GeV,
the oscillation is still
within its first cycle and can thus accommodate both the apparent surplus
of lower-energy mesons as well as the surplus of higher-energy baryons.
While there is therefore no apparent need for exotic mesons
in the mass range Freund and Rosner had in mind ($1.7 \leq M \leq 2$ GeV),
this oscillating ratio suggests an entirely different scenario for exotic
hadrons:  each repeating cycle of the oscillation may correspond
to the threshold for the next-order exotic mesons and baryons.
Other scenarios ({\it e.g.}, involving glueballs and hybrid
quark/gluon states) are possible as well.

Let us now be more specific, and first outline some of the basic results
of string theory (including that of Kutasov and Seiberg)
which will be relevant for our discussion.
Strings are one-dimensional extended objects whose
different vibrational and rotational configurations
correspond to different spacetime particles or states;
in general the mass of such a state is given by
\begin{equation}
        m~=~\sqrt{{n\over{\alpha'}}}~,~~~~~n\in {\bf Z}
\label{massrelation}
\end{equation}
where $\alpha'$ is a constant characterizing the energy scale of the theory
and where $n$ is related to the number of vibrational
mode-excitations necessary for producing the state.
Since the Lorentz spin $J$ of such a state must satisfy $J\leq n+\alpha_0$
where
$\alpha_0$ is a constant, we have the general result
\begin{equation}
      J ~\leq ~\alpha'\,m^2 ~+~ \alpha_0
\label{Regge}
\end{equation}
which identifies the constant $\alpha'$ as the traditional Regge slope.
If the particular string theory contains both bosonic and fermionic states,
we may denote their numbers at each level $n$ as $B_n$ and $F_n$
respectively;  note that these are the numbers of {\it states}
or field-theoretic degrees of freedom, and not the number of particles
({\it e.g.}, spin or isospin multiplets).
Another well-known prediction of string theory, then, is the asymptotic
exponential growth of these numbers as functions of $n$:
\begin{equation}
     B_n,~F_n ~\sim~ a \,n^{-b}\,e^{c\sqrt{n}} ~~~{\rm as} ~n\to\infty~
\label{hagedorn}
\end{equation}
where the positive constants $a$, $b$, and $c$ are theory-specific parameters.
Eqs.~(\ref{Regge}) and (\ref{hagedorn}) apply in general
to all string-type theories.
More recently, however, Kutasov and Seiberg have
obtained a result\cite{kutsei} which applies to those string theories
(or more generally, to those two-dimensional conformal field theories)
which are free of physical tachyons and which have modular-invariant
one-loop (toroidal) partition functions.
Specifically, if we define $B(N)\equiv \sum_{n=0}^N B_n$ and
$F(N)\equiv\sum_{n=0}^N F_n$, then Kutasov and Seiberg claim that
\begin{equation}
       \lim_{N\to \infty} ~[B(N) -F(N)] ~=~ 0~,
\label{strong}
\end{equation}
which in turn implies the weaker constraint
\begin{equation}
      \lim_{N\to\infty} ~[F(N)/B(N)] ~=~1.
\label{weak}
\end{equation}
We shall require only this weaker form of the Kutasov-Seiberg
result;  indeed, the stronger version in Eq.~(\ref{strong}) may not be
entirely correct.\cite{toappear}

The extent to which such a string theory can be taken as a theory
of hadrons is far from clear, and therefore in this letter we shall confine
ourselves
to only those issues which follow
from direct comparisons with the above generic results.
Specifically, we shall assume\cite{freuros}\ that
one can model hadronic physics as a GeV-scale string theory giving
rise to Eqs.~(\ref{Regge}), (\ref{hagedorn}), and (\ref{weak}),
with bosonic states identified as meson degrees of freedom
and fermionic states as baryon degrees of freedom;  furthermore,
we shall consider only those generic aspects of string theory
which affect the {\it relative}\/ numbers of these states ({\it i.e.}, their
ratio) or their separate {\it patterns}\/ of growth.
Any other features, such as the specific absolute sizes of $B(N)$ and
$F(N)$ or the mapping between particular string configurations
and particular hadronic states, are likely to be highly model-dependent.

We have computed the numbers and densities of experimentally-observed
meson and baryon states as functions of their masses.
We have included those states containing only
the three light quarks ($u,d,s$), both for reasons of experimental
statistics\cite{freuros}\
and more fundamentally because hadrons composed of heavy
quarks do not lie on linear Regge trajectories as a string picture
would dictate [Eq.~(\ref{Regge})].
We differ from Ref.~1, however, in recognizing that
although states in string theory are typically of zero width,
most of the hadronic states or resonances are quite broad.
Therefore, we have taken the hadronic density of states to be
a sum of normalized Breit-Wigner distributions:
\begin{equation}
         {{dN}\over{dm}} ~=~ {1 \over {2\pi}} \,\sum_i \, W_i \,
      {{\Gamma_i}\over{(m-M_i)^2+{\Gamma_i}^2/4}}
\label{Breit}
\end{equation}
where $M_i$ and $\Gamma_i$ are respectively the masses and widths of
the observed states,\cite{PPDB}
and where $W_i$ are their multiplicities [{\it i.e.}, the
number of {\it states} per resonance, or $(2I+1)(2J+1)$
for a charge self-conjugate state of spin $J$ and isospin $I$,
and twice that otherwise].
In Fig.~\ref{hagplot} we have plotted the total hadronic density of states
as a function of $m$, and it is clear that
this density experiences the exponential (Hagedorn-like) growth
suggested in Eq.~(\ref{hagedorn}) with Hagedorn temperature\cite{hagref}
$T_H\equiv (c\sqrt{\alpha'})^{-1}\approx 250$ MeV,
at least for masses up to 2 GeV.  Barring unexpected physics,
the failure of the curve in Fig.~\ref{hagplot} to maintain this growth
beyond 2 GeV is likely to be a reflection of
current experimental limitations.
Thus, we shall henceforth limit our attention to the
experimental data below 2 GeV.

In Fig.~\ref{BFplot} we have plotted the separate numbers
(or integrated densities) of baryon and
meson states with masses $m\leq M$ as functions of $M$.
In order to facilitate a comparison with Eq.~(\ref{weak}),
we have also plotted their {\it ratio} as the
shaded region in Fig.~\ref{stringplot}:
this shaded region indicates
the uncertainty in the ratio function due to the hadronic widths,
with the upper border of the region corresponding to the Breit-Wigner
densities in Eq.~(\ref{Breit})
and the lower border corresponding to the zero-width case.
Either way, several features are immediately apparent, among them
the pronounced surplus of mesons below 1.5 GeV and the pronounced
surplus of baryons above this energy;  indeed, this ratio
shows no sign of a plateau near unity.
This figure thus clearly indicates
that it is hardly compelling to interpret this mass region as the region of
onset of Kutasov-Seiberg asymptotic behavior.
It is in fact straightforward to estimate the string-level $n$ in
Eq.~(\ref{massrelation}) to which a mass of 1.5 GeV corresponds:  taking
the measured value of the hadronic Regge slope
$\alpha'\approx 0.9 ~(\rm GeV)^{-2}$, we obtain $n \approx 2$.
Indeed, the entire regions $< 2$ GeV
correspond only to string-levels $n\leq 4$.
Thus, even though these low-lying levels experience
the asymptotic growth in Eq.~(\ref{hagedorn}),
they clearly need not manifest the asymptotic behavior predicted in
Eq.~(\ref{weak});  indeed, the latter asymptotic behavior
occurs only at higher energies.

Therefore, in order to determine the characteristics of the {\it approach}\/
towards asymptotic behavior, we have calculated the ratio functions
$R(N)=F(N)/B(N)$ predicted by a variety of different string theories (or string
``models'') of the sort to which Eq.~(\ref{weak}) should apply.
While certain features of this function vary greatly and are
highly model-dependent, others -- such as the exponential increase
in the level degeneracies [Eq.~(\ref{hagedorn})] or the existence of a
Kutasov-Seiberg limit [Eq.~(\ref{weak})] -- indeed appear to be generic.
In particular, we find an important third universal
feature:\cite{toappear}~  as $N$ increases,
we find that the function $R(N)$ oscillates {\it around}\/
unity, with the amplitude of this oscillation decreasing with increasing
$N$.  This ``damped'' oscillation, periodic in $n=\alpha' M^2$,
is of course consistent with the Kutasov-Seiberg result in Eq.~(\ref{weak}).
Such an oscillation between bosonic and fermionic states is a
consequence (and in fact the signature) of an underlying string symmetry known
as modular invariance, and the wavelength $\lambda$
of this oscillation is determined only by the energy scale of the
theory,\cite{toappear}~ $\lambda=4/\alpha'$.
The amplitude, on the other hand, is somewhat model-dependent,
and in fact vanishes in the case of supersymmetry:  indeed,
the only way to break supersymmetry while preserving modular
invariance is to do so in this regular oscillatory manner.\cite{toappear}~
In Fig.~\ref{stringplot} we have superimposed the results
of a calculation based on a typical non-supersymmetric string model,
plotting $R(N)$ {\it vs.} $M\equiv \sqrt{N/\alpha'}$.

In the mass range $M\leq 2$ GeV,
the behavior of the string ratio in Fig.~\ref{stringplot}
is certainly consistent with
the observed ratio:  this oscillation typically
begins with $R<1$ (at $N=0$), first crosses $R=1$ at $N=2$
(corresponding to $M\approx 1.5$ GeV), and then increases
beyond 1 as $M$ approaches 2 GeV.   Thus we see that the sign of the
oscillation, as well as the position of the first node, are
consistent with the data, and a surplus of mesons below
1.5 GeV as well as a surplus of baryons above 1.5 GeV
are easily accommodated.
Thus, on the basis of a comparison between these two figures in the
$M\leq 2$ GeV range, we find that we need {\it not}\/ claim
a deficit of meson states with masses just above 1.5 GeV.

It will be interesting, however, to see whether the {\it entire}\/
string-theoretic oscillation is ultimately realized at higher energies.
While such an oscillation between bosonic and fermionic states has
not been observed experimentally,
we have seen in Fig.~\ref{hagplot} that
many hadronic states with energies above 2 GeV must be missing
if Hagedorn-like growth is to be maintained in that region.
That many such states are missing is also expected from
an SU(3) picture as well as from conventional Regge-trajectory arguments.
Such an oscillation, therefore, remains entirely possible.

It is important to bear in mind that we have focused on only the
generic features predicted by a generic string-type theory, and
one would need to further refine a particular string
picture in order to expect a more quantitative
agreement between the observed and predicted ratio functions.
For example, the string theories we have examined here are
intrinsically non-interacting:
all of their states (or particles) have zero width, and can populate
only the discrete energy levels indicated in Eq.~(\ref{massrelation}).
This is the origin of the sharp changes in the string ratio function in
Fig.~\ref{stringplot},
and a more fully-developed string theory incorporating particle interactions
would undoubtedly yield a smoother, more continuous ratio function.
Furthermore, dynamical considerations are also at the root of
the relatively small size of the experimentally observed ratio function
at masses $M\leq 1$ GeV: the lowest-lying mesons ({\it i.e.},
the pions) have masses protected by a nearly-unbroken
chiral symmetry, while the masses of the lowest-lying baryons
({\it i.e.}, the proton and neutron) are entirely unprotected
and consequently much greater.
This is in contrast to non-interacting string theories,
which generically contain both bosons {\it and}\/ fermions at the
(exactly) massless level.
A fully interacting string theory, therefore, should be expected
to yield a closer agreement between the ratio functions,
especially in the lower-mass region.
On the other hand, the {\it oscillations} in the ratio function
are of a more universal nature, and although interactions can be
expected to make them smooth, they should remain quite pronounced
in the region $M<4$ GeV where their amplitudes are large.

Given that string theories generically lead to such oscillations,
and given that we cannot soon expect to observe {\it all}\/ existing states
in the several-GeV region, it is natural to try to predict how
these oscillations might arise within the
context of a more traditional quark/gluon picture.  While the string
theories themselves unambiguously predict which string
vibrational/rotational configurations
are ultimately responsible for producing these oscillations,\cite{toappear}
one must specify or choose a particular mapping between these configurations
and the various
quark/gluon states in order to interpret these oscillations in terms of
selected groups of baryons and mesons.  The results are then highly
model-dependent.  Therefore, rather than advocate a particular
string-to-hadron mapping, we will simply propose two possible
resulting scenarios which naturally extend the ideas of Ref.~1.

One natural scheme which might lead to such a regular, periodic
meson/baryon oscillation involves exotic hadrons -- {\it i.e.},
mesons with quark structure $(q\overline{q})^{p+1}$
and baryons with quark structure $q^{p+3} \, \overline{q}^p$ for $p\geq 1$.
The special cases with $p=0$ of course correspond to the ordinary mesons and
baryons which respectively dominate the two halves of
the first cycle of the oscillation.
It is thus natural
to speculate that such a repeating pattern of oscillations is the
result of regularly-spaced thresholds for the $p^{\rm th}$ exotic hadrons,
implying alternating mass regions in
which either the $p^{\rm th}$ exotic mesons or baryons dominate:
\begin{eqnarray}
    (q\overline{q})^{p+1} ~{\rm mesons:}&~~~~(p+1/4)\,\lambda ~
     &\leq~M^2 ~\leq~ (p+1/2)\,\lambda~ \nonumber\\
    q^{p+3}\, \overline{q}^p~{\rm baryons:}&~~~~(p+3/4)\,\lambda ~
     &\leq~M^2 ~\leq~ (p+1)\,\lambda~
\label{exotics}
\end{eqnarray}
where $\lambda = 4/\alpha'\approx 4.4\, ({\rm GeV})^2$.
Such an ordering of thresholds is in fact consistent with
alternative analyses.\cite{freuros,margolis}
Another
scenario involves not only glueballs
but hadron/glue ``hybrids'', for such states --if color-neutral--
are in principle also present in a quark-gluon theory.
While glueballs are necessarily bosonic, hybrid states can
contribute to both bosonic and fermionic degrees of freedom depending
on their quark content.  In this scenario, then, each subsequent
cycle of our oscillation corresponds to the crossing of the threshold for
the next-order hybrid hadrons ({\it i.e.}, hadrons with one additional
gluonic insertion), with
the wavelength $\lambda=4/\alpha'$ of our oscillation
representing the mass shift resulting from such gluonic insertions.
Thus, this picture too can naturally explain the regularity of the
string-predicted oscillation.
Note, however, that {\it any} such picture necessarily implies the existence of
exponentially increasing numbers of fundamentally new hadronic
states at each of the mass regions listed in Eq.~(\ref{exotics}) --
starting with, in particular, several hundred between 2 and 2.3 GeV.

In summary, then, we find that a generic hadron-scale string
theory is consistent with the observed ratio of baryon and
meson states;  in particular, agreement with string theory does
not require the existence of ``missing mesons'' (ordinary or exotic)
in the mass region just above 1.5 GeV.
On the other hand, we find that string theory and modular
invariance predict a fermion/boson
ratio which {\it oscillates}\/ around unity as the mass increases,
with the amplitude of these oscillations steadily decreasing.
Such a picture therefore lends itself to a variety of
interpretations involving exotic and/or hybrid hadrons,
with each cycle of this oscillation corresponding to the thresholds
for the next-order mesons and baryons.
It will be interesting to see whether such pictures
can be realized in more traditional ({\it e.g.}, statistical
or potential) quark-models as well.

\bigskip
\bigskip
\bigskip
\centerline{\bf ACKNOWLEDGMENTS}
\bigskip
We are pleased to thank our colleagues at McGill, especially
C.S.~Lam and B.~Margolis, for many fruitful discussions.
This work was supported in part by NSERC (Canada)
and les fonds FCAR (Qu\'ebec).

\newpage
 ~~

\vskip 7.0truein

\figure{Total density of observed hadronic states as function of mass,
    along with best-fit to Hagedorn form of Ref.~2.\label{hagplot}}
\newpage
 ~~
\vskip 7.0truein

\figure{Total numbers of observed baryons (solid line) and mesons
    (dashed line) with masses $\leq M$, as functions of $M$.\label{BFplot}}
\newpage
 ~~
\vskip 7.0truein

\figure{{\it Shaded region}:  observed ratio of numbers of baryon and
    mesons, as discussed in text.
    {\it Solid line}:   ratio function from a typical string
    model.\label{stringplot}}

\eject

\end{document}